\documentclass[aps,prb,twocolumn,showpacs,groupedaddress,preprintnumbers,amsmath,amssymb]{revtex4}
\usepackage{amsmath,amssymb,epsfig,color,pstricks,bm,alltt,graphicx}

\usepackage[colorlinks,plainpages=false,linkcolor=black,urlcolor=blue,citecolor=black,pdfpagemode=UseNone,pdfstartview=FitBH]{hyperref}

\definecolor{MyDarkBlue}{rgb}{0,  0.3,  0.9}
\definecolor{MyDarkBlack}{rgb}{0,  0,  0}
\newcommand \modified[1]{\textcolor{MyDarkBlack}{#1}}

\definecolor{darkgreen}{rgb}{0,0.5,0}

\begin{document}
\title
{ARPES spectral functions and Fermi surface for La$_{1.86}$Sr$_{0.14}$CuO$_4$
compared with LDA+DMFT+$\Sigma_{\textbf{k}}$ calculations}
\author{I.A. Nekrasov$^1$, E.E. Kokorina$^1$, E.Z. Kuchinskii$^1$, M.V. Sadovskii$^1$,\\
S. Kasai$^3$, A. Sekiyama$^3$, S. Suga$^3$}

\affiliation
{$^1$Institute for Electrophysics, Russian Academy of Sciences,
Ekaterinburg, 620016, Russia\\
$^3$Graduate School of Engineering Science, Osaka University, Toyonaka,
Osaka 560-8531, Japan}

\date{\today}

\begin{abstract}

Slightly underdoped high-T$_c$ system La$_{1.86}$Sr$_{0.14}$CuO$_4$ (LSCO) is studied by means
of high energy high resolution angular resolved photoemission spectroscopy (ARPES) and combined
computational scheme LDA+DMFT+$\Sigma_{\textbf{k}}$. Corresponding one band Hubbard model is
solved via dynamical mean-field theory (DMFT), while model parameters needed are obtained from
first principles within local density approximation (LDA). 
An ``external'' ${\bf k}$-dependent self-energy $\Sigma_{\textbf{k}}$
describes interaction of correlated electrons with antiferromagnetic (AFM) pseudogap fluctuations.
Experimental and theoretical data clearly show ``destruction'' of the LSCO Fermi surface
in the vicinity of the ($\pi$,0) point and formation of ``Fermi arcs'' in the nodal directions.
ARPES energy distribution curves (EDC) as well as
momentum distribution curves (MDC) demonstrate deviation of the quasiparticle band from
the Fermi level around ($\pi$,0) point. The same behavior of spectral functions
follows from theoretical calculations suggesting AFM origin of the pseudogap state.

\end{abstract}


\maketitle

\section{Introduction}

One of the not yet solved puzzles of cuprate
high-temperature superconductors (HTSC) is a nature of
underdoped normal state - the pseudogap regime\cite{pseudogap}.
Perhaps most powerful experimental tool to access electronic
properties of the pseudogap state is angular resolved photoemission
spectroscopy (ARPES)\cite{ARPES, ARPES1,ARPESYoshida2006}.
Common understanding is the fluctuating origin of the pseudogap state,
however type of the fluctuations is still under discussion.
Whether these are superconducting fluctuations\cite{fluctuations}
or some order parameter fluctuations (AFM(SDW), CDW, stripes, etc.)\cite{Bi2212, nonmonPG}
coexisting or competing with
Cooper pairing is up to now undecided.

There are several prototype compounds among high-T$_c$ systems e.g.
hole doped Bi$_2$Sr$_2$CaCu$_2$O$_{8-\delta}$ (Bi2212) system or
the electron doped one --
Nd$_{2-x}$Ce$_x$CuO$_4$ (NCCO). Now a lot of experimental 
ARPES data on Bi2212 and NCCO is available (see the review~Ref.~\onlinecite{ARPES}).
For instance Fermi surface (FS) maps, quasiparticle band dispersions and
even self-energy lineshapes  mapped onto some models 
are obtained from modern ARPES measurements~\cite{nonmonPG}.
Into this list of prototype componds should be of course included
the first ever high-T$_c$ hole doped system La$_{2-x}$Sr$_{x}$CuO$_4$ (LSCO)
which was also investigated both in theory and experiment with great details.\cite{ARPES}

Within the normal underdoped phase (pseudogap regime)
a number of interesting physical phenomena were discovered.
For example, FS is partially ``destructed'' in the vicinity of the so-called
``hot-spots'' (points of crossing between FS and AFM umklapp surface).
``Shadow bands'' (partial folding of band dispersion)
appear apparently as a result of short range AFM order.
Formation of the so called Fermi ``arcs'' around Brillouin zone (BZ) diagonal
reminiscent of the parts of noninteracting FS is experimentally detected
in numerous ARPES experiments\cite{ARPES}.
Despite apparently the same physics behind, pseudogap regime demonstrates
some material specific features. For Bi2212 Fermi ``arcs'' go almost up to
BZ border where they are strongly blurred.
The NCCO also has Fermi ``arcs'' but FS ``destruction'' looks different.
The ''hot-spots'' are well observed in NCCO while towards the BZ border
FS is almost restored as a non interacting one.\cite{NCCO_work}

The present paper is devoted to pseudogap behavior in the underdoped LSCO
and its comparison with Bi2212 and NCCO.

According to common knowledge high-T$_c$ systems are usually doped Mott insulators,
effectively described by the Hubbard model. Most common method
to solve the Hubbard model in our days is dynamical mean-field theory (DMFT)\cite{DMFT_method}.
Its exactness in the infinite spatial dimensions limit makes DMFT a local approach.
Since high-T$_c$ compounds as it is well established have quasi two-dimensional nature
spatial fluctuations play important role for their physics.
To overcome this difficulty we introduced
DMFT+$\Sigma_{\bf k}$ computational scheme\cite{fsdistr,dmftsk,FNT}
which supplies conventional DMFT with ``external''  ${\bf k}$-dependent
self-energy. Main assumption of the DMFT+$\Sigma_{\bf k}$ is additive form of the self-energy
which allows one to keep conventional DMFT self-consistent set of equations.
The DMFT+$\Sigma_{\bf k}$ approach was used to address
the pseudogap problem\cite{dmftsk}, electron-phonon
coupling in strongly correlated systems\cite{ephdsk} and disorder induced
metal-insulator transition within the Hubbard-Anderson model\cite{ham}.
For the pseudogap state this self-energy $\Sigma_{\bf k}$ describes the
interaction of correlated electrons with non-local (quasi) static short-ranged collective Heisenberg-like 
antiferromagnetic (AFM or SDW-like) spin fluctuations\cite{AFM_fluctuations}.
DMFT+$\Sigma_{\bf k}$ approximation was also shown to be appropriate to describe
two-particle properties e.g. optical conductivity.\cite{opt}

As a possible way of theoretical simulation of the pseudogap regime for real materials
we proposed LDA+DMFT+$\Sigma_{\textbf{k}}$ hybrid method\cite{Bi2212}.
It combines  first principle one-electron density functional theory calculations within
local density approximation (DFT/LDA) \cite{DFT_LDA} with DMFT+$\Sigma_{\textbf{k}}$.\cite{psik}

LDA+DMFT+$\Sigma_{\textbf{k}}$ method allowed us to obtain
Fermi arcs and ``hot-spots'' behavior for both electron doped e.g.
Nd$_{1.85}$Ce$_{0.15}$CuO$_4$\cite{NCCO_work} (NCCO) and
Pr$_{1.85}$Ce$_{0.15}$CuO$_4$ (PCCO)\cite{PCCO_work} as well as hole doped
Bi$_2$Sr$_2$CaCu$_2$O$_{8-\delta}$ (Bi2212)\cite{Bi2212} high-T$_c$ cuprates.
Pseudogap behavior of dynamic optical conductivity within LDA+DMFT+$\Sigma_{\textbf{k}}$\cite{opt}
was also discussed for B12212\cite{Bi2212} and NCCO\cite{NCCO_work}.

This communication reports LDA+DMFT+$\Sigma_{\textbf{k}}$ computations of
Fermi surface and spectral functions for hole underdoped
La$_{1.86}$Sr$_{0.14}$CuO$_4$ (LSCO) system supported by
high energy, high resolution bulk sensitive ARPES\cite{ARPES1}.

\section{Computational details}

The La$_2$CuO$_4$ system has base-centered orthorhombic crystal structure
with space group $Bmab$ with two formula units per cell\cite{Crystal_structure}.
Corresponding lattice parameters are a=5.3346, a=5.4148 and c=13.1172 \AA.
Atomic positions are the following: La(0.0,-0.0083,0.3616),
Cu(0,0,0), O(1/4,1/4,-0.0084), O2(0.0,0.0404,0.1837).

As a first step of LDA+DMFT+$\Sigma_{\textbf{k}}$ method
we performed density functional theory
calculations in the local density approximation (LDA)
for these crystallographic data. Band structure
was obtained within the linearized muffin-tin orbitals (LMTO) method\cite{LMTO}.
It is well known for these compounds that Fermi level is crossed by
antibonding O2$p$-Cu3$d$ partially filled orbital of $x^2-y^2$ symmetry.
Tight-binding parameters for this band were calculated by
\modified{$N$}-th order LMTO (NMTO) method\cite{NMTO} as (in eV units)
$t=-0.476$, $t'=0.077$, $t''=-0.025$, $t'''=-0.015$.
These values agree well with previous studies.\cite{hops}
Coulomb interaction value on effective Cu-3$d$($x^2-y^2$) orbital was calculated by
constrained LDA approach\cite{Gunnarsson} and was found to be $U$=1.1~eV.
These LDA obtained parameters are used to set up corresponding one-band Hubbard model.

The second step is further treatment of the above defined Hubbard model
within the dynamical mean-field theory (DMFT) self-consistent set of equations\cite{DMFT_method}
supplied by ``external'' momentum-dependent self-energy
$\Sigma_{\textbf{k}}$\cite{dmftsk}.
Using the additive form of self-energy (main approximation of the scheme 
neglecting the interference between Hubbard interaction and pseudogap fluctuations,
which allows one to preserve conventional DMFT equations)
one can define LDA+DMFT+$\Sigma_{\textbf{k}}$ Green function as:
\begin{equation}
G_{\bf k}(\omega)=\frac{1}{\omega+\mu-\varepsilon({\bf k})-\Sigma(\omega)
-\Sigma_{\bf k}(\omega)}
\label{Gk}
\end{equation}
where bare electron dispersion $\varepsilon({\bf k})$ is defined by LDA calculated hopping parameters
listed above. To calculate $\Sigma_{\textbf{k}}$ we used a two-dimensional pseudogap
model\cite{pseudogap,AFM_fluctuations} describing nonlocal correlations induced
by (quasi) static short-range collective Heisenberg-like AFM spin fluctuations.
Thus we introduce correlation length dependence of the pseudogap fluctuations into conventional DMFT loop.

There are two points which make DMFT+$\Sigma_{\textbf{k}}$ different from the usual DMFT.
First, momentum dependent $\Sigma_{\textbf{k}}$ is recalculated on each DMFT iteration
($\Sigma_{\bf k}(\mu,\omega,[\Sigma(\omega)])$ is in fact the function of DMFT chemical potential and
DMFT self-energy). Second, DMFT+$\Sigma_{\textbf{k}}$ lattice problem is defined 
at each DMFT iteration as:
\begin{equation}
G_{ii}(\omega)=\frac{1}{N}\sum_{\bf k}\frac{1}{\omega+\mu
-\varepsilon({\bf k})-\Sigma(\omega)-\Sigma_{\bf k}(\omega)}.
\label{Gloc}
\end{equation}
After numerical self-consistency is reached we get the Green's function (\ref{Gk})
with corresponding $\Sigma(\omega)$ and $\Sigma_{\bf k}(\omega)$ taken on the last DMFT iteration.
All further computational details can be found e.g. in Refs.~\onlinecite{Bi2212,dmftsk,NCCO_work}.

As an ``impurity solver'' for DMFT equations
numerical renormalization group (NRG\cite{NRG,BPH}) was employed.
Temperature of DMFT(NRG) computations was taken to be 0.011~eV and
electron concentration used was n=0.86.

Self-energy $\Sigma_{\textbf{k}}(\omega)$ due to pseudogap fluctuations depends, in general,
on two parameters: the pseudogap amplitude $\Delta$, 
and the correlation length $\xi$.\cite{pseudogap,AFM_fluctuations}
The value of $\Delta$ was calculated as in Ref.~\onlinecite{dmftsk}
\begin{eqnarray}
\Delta^2=
U^2\frac{<n_{i\uparrow}n_{i\downarrow}>}{n^2}<(n_{i\uparrow}-n_{i\downarrow})^2>,
\label{DeltHubb}
\end{eqnarray}
where local densities $n_{i\uparrow},~n_{i\downarrow}$ and  double occupancy $<n_{i\uparrow}n_{i\downarrow}>$
were calculated within the standard DMFT\cite{DMFT_method}.
Behavior of $\Delta$ as a function of hopping integrals and Coulomb interaction was studied in our previous work
\cite{dmftsk}, while $\Delta$ as a function of occupancy $n$ was investigated in
Ref.~\onlinecite{Bi2212}.
For $\xi$ we believe it is more safe to take experimental values.
In this work the value of $\Delta$ was calculated to be 0.275~eV and 
$\xi$ was taken to be 10$a$, where $a$ - lattice constant.\cite{LSCO_xi}

\section{Experimental details}

The high-energy ARPES measurements were carried out at BL25SU in SPring-
8, using incident photons of 500 eV, on single crystal samples.
The normal to the cleaved sample surface was set almost parallel to
the axis of the lens of the analyzer and the sample was set to about 45$^\circ$ from the incident light 
direction. The photoelectrons within polar angles of about $\pm6^\circ$ around
the normal to the sample were simultaneously collected using a GAMMADATASCIENTA
SES200 analyzer, thereby covering more than a whole Brillouin zone along the
directions of the analyzer slit. The Fermi surface mapping was performed by changing the angle
along the perpendicular direction to the analyzer slit. The base pressure was about 4 x
10$^{-8}$Pa. The (001) clean surface was obtained by cleaving the samples in situ in vacuum
at the measuring temperature of 20 K. The overall energy resolution was set to 100 and
170 meV for high-resolution measurements and Fermi surface mapping, respectively.
The angular resolution was $\pm$0.1 ($\pm$0.15) for the perpendicular (parallel) direction to
the analyzer slit. These values correspond to the momentum resolution of $\pm$0.024$\pi/a$ (
$\pm$0.036$\pi/a$) at h$\nu$=500 eV, 
where $a$ is twice the Cu-O bondlength within the CuO$_2$ plane. 
By virtue of the longer photoelectron mean free path of $\sim$12 {\AA} at the kinetic energy 
of $\sim$500 eV than that for conventional ARPES at $h\nu\sim$20-60 eV, 
the bulk contribution to the spectral weight is estimated as about 60\%.
The position of the Fermi level was calibrated with Pd spectra.

\section{Results and discussion}

\begin{figure}[ht]

\includegraphics[width=0.85\columnwidth]{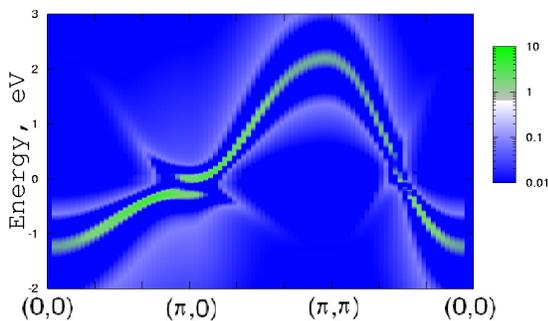}
\caption{LCOO Cu-3$d$($x^2-y^2$) band dispersion 
along high symmetry directions of square Brillouin zone
computed with LDA+DMFT+$\Sigma_k$.
The Fermi level is zero.
\label{Contour_plot}}
\end{figure}

In case of finite temperature and interaction strength we have to take into account the
finite life time effects of quasiparticles. Thus instead of just a dispersions $\varepsilon({\bf k})$
one should rather work with the spectral function  $A(\omega,\textbf{k})$ given by
\begin{equation}
A(\omega,\textbf{k})=-\frac{1}{\pi}{\rm Im} G(\omega,\textbf{k}),
\label{specf}
\end{equation}
with retarded Green's function $G(\omega,\textbf{k})$ obtained via LDA+DMFT+$\Sigma_{\textbf{k}}$
scheme\cite{fsdistr,dmftsk,opt}. Of course there are
considerable lifetime effects originating from the $\Sigma_{\bf k}$
corresponding to interaction with AFM fluctuations
(substituted in our approach by the quenched random field).

In Fig.~\ref{Contour_plot} contour map
of spectral function (\ref{specf}) obtained from LDA+DMFT+$\Sigma_{\textbf{k}}$
for Cu-3$d$($x^2-y^2$) band is presented. The width of the spectral function is
inversely proportional to the lifetime. Around ($\pi$,0) point one can clearly
see splitting of the spectra by AFM pseudogap fluctuations of the order of 2$\Delta$. Also
AFM nature of the pseudogap fluctuations leads to formation of the ``shadow'' band
which is much weaker in intensity and becoming the real quasiparticle band 
in case of complete folding in case of long-range AFM order. 

\begin{figure}[t]

\includegraphics[width=0.85\columnwidth]{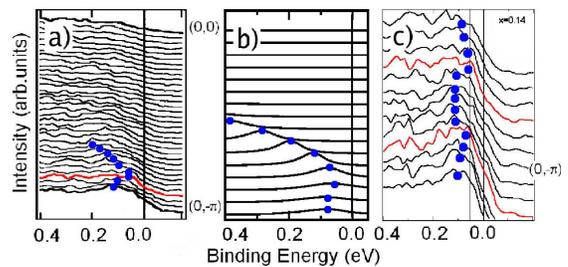}
\caption{ARPES EDC curves (a) and LDA+DMFT+$\Sigma_k$ spectral functions (b)
along (0,0)-($\pi,0$) high symmetry direction; (c) ARPES MDC curves around ($\pi,0$) point
for LSCO at x=0.14.
On panels (a), (b) and (c) filled circles guide the motion of the $A(\omega,{\bf k})$ maxima.
The Fermi level is zero.
\label{sf}}
\end{figure}

Figure~\ref{sf} displays experimental energy distribution curves (EDC) on the panel (a) along
(0,0)--($\pi$,0) direction. Around ($\pi$,0)-point as shown by stars certain deviation of the 
$A(\omega,\textbf{k})$ maxima from the Fermi level a kind of ``turn-back'' is observed.
We attribute such behavior of the $A(\omega,\textbf{k})$
to pseudogap fluctuations. Similar theoretical behavior is shown on panel (b)\cite{shift} of Fig.~\ref{sf}
as calculated by our LDA+DMFT+$\Sigma_{\textbf{k}}$ approach (see also Fig.~\ref{Contour_plot}).
The same behavior is also observed as traced by circles in experimental ARPES momentum distribution curves (MDC)
demonstrated on panel (c) of Fig.~\ref{sf}.

The bulk-sensitive high-energy ARPES data for La$_{1.86}$Sr$_{0.14}$CuO$_{4}$ show 
a clear ``turn-back'' structure of the 
EDC peak as a function of momentum near (0,$-\pi$), which were not seen in  the 
previous low-energy ARPES data 
for La$_{1.85}$Sr$_{0.15}$CuO$_{4}$ ~\cite{ARPESYoshida2006}. 
The contour map of the spectral weight in the vicinity of $E_F$ seems to 
be essentially similar in overall features for this doping level between the high-energy 
and low-energy ARPES studies. 

\begin{figure}[b]
\includegraphics[clip=true,width=1.\columnwidth]{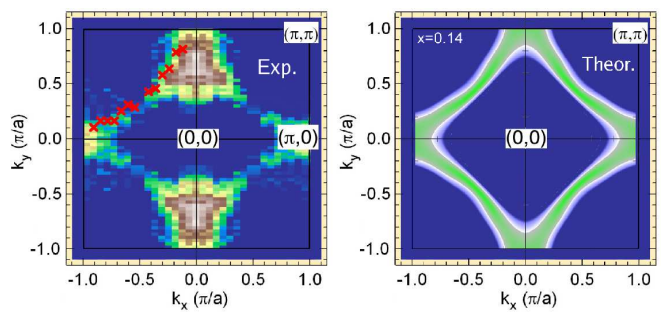}\\
\caption{ Fermi surfaces of LSCO at x=0.14 from experiment (left panel) and LDA+DMFT+$\Sigma_{\textbf{k}}$
computations (right panel) Red crosses on the left panel correspond to experimental ${\bf k}_F$ values.
\label{FermiSurface}}
\end{figure}

Experimental and theoretical Fermi surface maps are shown in Fig.~\ref{FermiSurface}
at panels (a) and (b) correspondingly. Both pictures reveal strong scattering around
($\pi$,0)-point which we associate with scattering in the vicinity of the so-called ``hot-spots''
(cross-points of a Fermi and AFM umklapp surfaces) which are close to the ($\pi$,0)\cite{Bi2212,NCCO_work}.
Such strong scattering comes from scattering processes with momentum transfer of the order of
{\bf Q}=($\pi,\pi$)\cite{pseudogap,AFM_fluctuations},
corresponding to AFM pseudogap fluctuations. Along nodal directions
we observe typical Fermi arcs. They are pretty well seen in the theoretical data
while for experiment we have just narrow traces of them.

\section{Conclusion}

Our LDA+DMFT+$\Sigma_{\textbf{k}}$ hybrid approach
was shown to be an effective numerical tool to describe short-range ordered
state in quasi-two dimensional systems.\cite{Bi2212,NCCO_work,PCCO_work}
Material specific model parameters such as hopping integrals (which define bare
electronic band dispersion of effective Cu$3d(x^2-y^2)$ orbital) were calculated
via LDA based NMTO method.\cite{NMTO} Coulomb interaction parameter $U$ was obtained from constrained LDA method.
Pseudogap amplitude $\Delta$ was computed using LDA+DMFT scheme.\cite{dmftsk,NCCO_work}
Supplementing conventional DMFT self-energy by $\Sigma_{\textbf{k}}(\omega)$
describes non-local dynamical correlations due to  short ranged
collective Heisenberg like AFM spin fluctuations.

In this work we performed LDA+DMFT+$\Sigma_{\textbf{k}}$ calculations
for hole-doped La$_{1.84}$Sr$_{0.14}$CuO$_4$ compound in the pseudogap regime.
Because of fluctuations of AFM
short-range order we clearly observe formation of the so called ``shadow bands''
as partially folded bare dispersion. Pseudogap is formed around ($\pi$,0)
point which is qualitatively the same as in Bi2212\cite{Bi2212}, NCCO\cite{NCCO_work}
and PCCO\cite{PCCO_work}. As to Fermi surface of LSCO it is alike that obtained
for Bi2212\cite{Bi2212}. Namely the ``hot spots'' are not well resolved since
the crossing point of the bare Fermi surface and AFM umklapp surface are very close to
Brillouin zone border.  This fact is essentially due to
the shape and size of LDA Fermi surface. In this respect, situation here is
different from that for NCCO\cite{NCCO_work}
and PCCO\cite{PCCO_work}, where ``hot spots'' are clearly seen.
To support these theoretical results we present here the new high energy, high resolution ARPES data for LSCO.
Typical pseudogap-like effects of Fermi surface destruction were observed in both theory
and experiment. The same is true for spectral functions.
The overall semiquantitative agreement between theory and experiments
basically supports our general picture of the pseudogap state as due to
strong scattering of carriers by short-range AFM order fluctuations.

\section{Acknowledgements}

We thank Thomas Pruschke for providing us the NRG code. This work is supported by
RFBR grants 08-02-00021, 08-02-91200, RAS programs
``Quantum physics of condensed matter'' and ``Strongly correlated electrons solids''.
IN is supported by Grants of President of  MK-614.2009.2(IN) and
Russian Science Support Foundation.
AS and SS acknowledge JSPS for the financial support for the Japan-Russia collaborating research
in 2008-09. They also acknowledge MEXT, Japan for the support in Grant-in-Aid for scientific
Research (18104007, 15GS0123, 18684015).



\begin{thebibliography}{99}
\raggedright

\bibitem{pseudogap} T. Timusk, B. Statt, Rep. Progr. \modified{Phys.}, {\bf 62}, 61 (1999);
M.\,V.\,Sadovskii, Usp. Fiz. Nauk {\bf 171} 539 (2001) [Phys. Usp. {\bf 44}, 515 (2001)];
M.\,V.\,Sadovskii, in ``Strings, branes, lattices, networks, pseudogaps and dust'',
Scientific World, Moscow, 2007, p. 357 (in Russian), English version:
cond-mat/0408489.

\bibitem{ARPES}
A.\,Damascelli, Z.\,Hussain, and Zhi-Xun\,Shen\modified{,}  Rev. Mod. Phys. \textbf{75}, 473 (2003).

\bibitem{ARPES1} M. Tsunekawa, A. Sekiyama, S. Kasai, S. Imada, H. Fujiwara,
T. Muro, Y. Onose, Y. Tokura and S. Suga,  New J. Phys. \textbf{10}, 073005 (2008).

\bibitem{ARPESYoshida2006}T. Yoshida, X. J. Zhou, K. Tanaka, W. L. Yang, Z. Hussain, 
Z.-X. Shen, A. Fujimori, S. Sahrakorpi, M. Lindroos, R. S. Markiewicz, A. Bansil, 
Seiki Komiya, Yoichi Ando, H. Eisaki, T. Kakeshita, and S. Uchida, 
Phys. Rev. B \textbf{74}, 224510 (2006). 

\bibitem{fluctuations} \modified{V. J. Emery, S. A. Kivelson, Nature \textbf{374}, 434 (2002).}

\bibitem{nonmonPG} \modified{A. A. Kordyuk, S.\,V.\,Borisenko, V.\,B.\,Zabolotnyy, R.\,Schuster, D.\,S.\,Inosov, D.\,V.\,Evtushinsky, A.\,I.\,Plyushchay, R. Follath, A. Varykhalov, L. Patthey, and H. Berger, Phys. Rev. B \textbf{79}, 020504(R) (2009).}

\bibitem{Bi2212} E.\,Z.\,Kuchinskii, I.\,A.\,Nekrasov, Z.\,I.\,Pchelkina, M.\,V.\,Sadovskii,
Zh. Eksp. Teor. Fiz. {\bf 131}, 908 (2007) [JETP {\bf 104}, 792 (2007)];
I.\,A.\,Nekrasov, E.\,Z.\,Kuchinskii, Z.\,V.\,Pchelkina and M.\,V.\,Sadovskii,
Physica C {\bf 460-462}, 997 (2007).

\bibitem{NCCO_work} E.\,E.\,Kokorina, E.\,Z.\,Kuchinskii, I.\,A.\,Nekrasov, Z.\,V.\,Pchelkina, M.\,V.\,Sadovskii, A.\,Sekiyama,
S.\,Suga, M.\,Tsunekawa. Zh. Eksp. Teor. Fiz. {\bf 134}, 968 (2008) [JETP {\bf 107}, 818 (2008)];
I.\,A.\,Nekrasov {\it et al.},
J. Phys. Chem. Solids {\bf 69}, 3269 (2008).

\bibitem{DMFT_method} A.\,Georges, G.\,Kotliar, W.\,Krauth and M.\,J.\,Rozenberg,
Rev. Mod. Phys. {\bf 68}, 13 (1996).

\bibitem{fsdistr}
E.\,Z.\,Kuchinskii, I.\,A.\,Nekrasov, M.\,V.\,Sadovskii,
Pis'ma ZhETF {\bf 82}, 217 (2005) [JETP Lett. {\bf 82}, 198 (2005)].

\bibitem{dmftsk}
M.\,V.\,Sadovskii, I.\,A.\,Nekrasov, E.\,Z.\,Kuchinskii, Th. Prushke,
V.\,I.\,Anisimov. Phys. Rev. B {\bf 72}, 155105 (2005).

\bibitem{FNT} E.Z. Kuchinskii, I.A. Nekrasov, M.V. Sadovskii,
Fizika Nizkikh Temperatur {\bf 32} (2006) 528
[Low Temperature Physics {\bf 32} (2006) 398].

\bibitem{ephdsk} E.Z. Kuchinskii, I.A. Nekrasov, M.V. Sadovskii, Phys. Rev. B {\bf 80}, 115124 (2009).

\bibitem{ham} E.Z. Kuchinskii, I.A. Nekrasov, M.V. Sadovskii,
Zh. Eksp. Teor. Fiz. {\bf 133}, 670 (2008) [JETP {\bf 106}, 581 (2008)].

\bibitem{AFM_fluctuations} J. Schmalian, D. Pines, B.Stojkovic, Phys. Rev. B {\bf 60}, 667 (1999);
E.Z. Kuchinskii, M.V. Sadovskii, Zh. Eksp. Teor. Fiz. {\bf 115},
1765 (1999) [JETP {\bf 88}, 347 (1999)].

\bibitem{opt}
E.\,Z.\,Kuchinskii, I.\,A.\,Nekrasov, M.\,V.\,Sadovskii, Phys. Rev. B {\bf 75} 115102 (2007).

\bibitem{DFT_LDA}  R. O. Jones and O. Gunnarsson, Rev. Mod. Phys. {\bf 61}, 689 (1989).

\bibitem{psik}K. Held, I. A. Nekrasov, G. Keller, V. Eyert, N. Blumer,
A. K. McMahan, R. T. Scalettar, Th. Pruschke, V. I. Anisimov, and D.
Vollhardt, Psi-k Newsletter {\bf 56}, 65 (2003)
[psi-k.dl.ac.uk/newsletters/News\_56/Highlight\_56.pdf];
K. Held, Adv. Phys. {\bf 56}, 829 (2007)

\bibitem{PCCO_work} I.A. Nekrasov, N.S. Pavlov, E.Z. Kuchinskii, M.V. Sadovskii, Z.V. Pchelkina, V.B. Zabolotnyy, J. Geck, B.Buchner, S.V. Borisenko, D.S. Inosov, A.A. Kordyuk, M. Lambacher, A. Erb,
Phys. Rev. B {\bf 80} 115124(R) (2009).

\bibitem{Crystal_structure} M. Braden, P. Schweiss, G. Heger, W. Reichardt, Z. Fisk,
K. Gamayunov, I. Tanaka, H. Kojima, Physica C {\bf 223}, 396 (1994).

\bibitem{LMTO} O. K. \modified{Andersen} , Phys. Rev. B {\bf 12}, 3060 (1975);
O. K. Andersen and O. Jepsen, Phys. Rev. Lett. {\bf 53}, 2571 (1984).

\bibitem{NMTO}O.\,K.\,Andersen and T.\,Saha-Dasgupta, Phys. Rev. B {\bf 62}, R16219 (2000);
O.\,K.\,Andersen {\it et al.}
Psi-k Newsletter {\bf 45}, 86 (2001); O.\,K.\,Andersen, T.\,Saha-Dasgupta, S.\,Ezhov, Bull. Mater. Sci. {\bf 26}, 19 (2003).

\bibitem{hops}Korshunov M.M., Gavrichkov V.A., Ovchinnikov S.G., Pchelkina Z.V.,
Nekrasov I.A., Korotin M.A., Anisimov V.I.,
Zh. Eksp. Teor. Fiz. {\bf 126}, 642 (2004) [JETP {\bf 99}, 559 (2004)].

\bibitem{Gunnarsson} O.\,Gunnarsson, O.\,K.\,Andersen, O.\,Jepsen, and J.\,Zaanen,
Phys. Rev. B \textbf{39}, 1708 (1989).

\bibitem{NRG} K.\,G.\,Wilson,  Rev.\ Mod.\ Phys. {\bf 47}, 773 (1975);
  H.\,R.\,Krishna-murthy, J.\,W.\,Wilkins, and K.\,G.\,Wilson,
  Phys.\ Rev.\ B {\bf 21}, 1003 (1980); {\it ibid.} {\bf 21}, 1044 (1980).
\bibitem{BPH}R.\,Bulla,  A.\,C.\,Hewson and Th.\,Pruschke,
  J.\,Phys.:\,Condens.\,Matter {\bf 10}, 8365 (1998);

\bibitem{LSCO_xi}M. Hücker, Young-June Kim, G. D. Gu, J. M. Tranquada, B. D. Gaulin,
J. W. Lynn, Phys. Rev. B {\bf 71}, 094510 (2005).


\bibitem{shift} Theoretical curves are shifted up by 0.2 eV for better agreement with experiment.


\end{thebibliography}
\end{document}